\documentclass[hyper,letterpaper,11pt]{JHEP3}
\usepackage{amsmath,amssymb,multirow,array}
\usepackage{graphicx}
\usepackage{cite}

\newcommand\beq{\begin{equation}}
\newcommand\eeq{\end{equation}}
\newcommand\be{\begin{equation}}
\newcommand\ee{\end{equation}}

\title{Conductivities for Hyperscaling Violating Geometries}

\preprint{\today}

\author{Andreas Karch\\
Department of Physics, University of Washington, Seattle, WA 98195, USA \\
Email: akarch@uw.edu
}

\abstract{We show that many results about holographic conductivities in geometries with hyperscaling violating scaling can be reproduced from simple scaling laws in the dual field theory. We show that the electro-magnetic response of probe branes in these systems require at least one additional scaling parameter $\Phi$ beyond the usual dynamical exponent $z$ and hyperscaling violating exponent $\theta$, as also pointed out in earlier work. We show that the scaling exponents can be chosen in such a way that the temperature dependence of DC conductivity and Hall angle in strange metals can be reproduced.
}

\begin{document}
\tableofcontents


\section{Introduction}

Holography \cite{Maldacena:1997re,Witten:1998qj,Gubser:1998bc} has recently been used to construct putative novel phases of compressible matter. While many examples of theories with holographic dual are known, the simplest holographic duals correspond to scale invariant theories. The original examples of holography describe conformally invariant relativistic systems in terms of Einstein gravity on anti-de Sitter (AdS) space. These have been generalized to backgrounds incorporating a non-trivial dynamical critical exponent $z$ \cite{Kachru:2008yh} as well as hyperscaling violating (HV) exponent $\theta$ \cite{Ogawa:2011bz,Huijse:2011ef,Dong:2012se}. For $\theta=0$ the holographically calculated conductivity, and in particular its dependence on temperature, can be understood in the field theory from a few very basic scaling laws \cite{Hartnoll:2009ns}: the dynamical critical exponent is defined via the dispersion relation
\beq \omega \sim k^z. \eeq
This forces us to assign scaling dimensions
\beq [x]=-1, \quad [t]=-z \eeq
to space and time. In an equilibrium system this fixes also the scaling of temperature
\beq [T] = z .\eeq
These relations still are true in the presence of hyperscaling violation; the latter only manifests itself in an anomalous scaling of the energy and free energy density
\beq \label{free} [\epsilon]=[f] = d+z-\theta \eeq
as well as the quantities that are derived from variations of $f$. In particular, one finds for the entropy density
\beq [s] = [f]-[T] = (d-\theta) \quad \Rightarrow \quad s\sim T^{\frac{d-\theta}{z}}. \eeq
Generalizing this scaling to the electro-magnetic sector one naively would postulate for vector potentials, chemical potential, electric and magnetic field as well as charge density and current
\beq
\label{naturalscalings}
[A_i]=1, \quad [A_0]=[\mu]=z, \quad [E]=1+z, \quad [B]=2, \quad [n]=d-\theta, \quad [j]=d-\theta+z-1 .\eeq
We will however see that these naive scalings are inconsistent with the holographic results for DC conductivities. In order to reproduce the properties of gauge fields in HV geometries we need to allow for anomalous scalings of the gauge fields\footnote{A somewhat similar proposal has been recently made in \cite{Khveshchenko:2014nka}. This reference tried to introduce the anomalous scaling into $[x]$ and $[t]$ directly. While this can correctly reproduce some of the holographic conductivity results, it does not lead to a self-consistent assignment for several reasons. Maybe most importantly, the scalings of \cite{Khveshchenko:2014nka} are inconsistent with the basic dispersion relation $\omega \sim k^z$. More concretely, in order to reproduce the correct conductivities we determine in here one would need the version of the scalings in \cite{Khveshchenko:2014nka} which symmetrically distributes the anomalous scalings between $[t]$ and $[x]$. In that case one obtains incorrect thermodynamics. Last but not least, we can derive our scaling from the properties of the holographic dual directly.}
\beq [A_i]=1 - \Phi, \quad [A_0]=z - \Phi .\eeq
Unlike $z$ and $\theta$, $\Phi$ not only depends on the bulk geometry but encodes properties of the bulk gauge field action. For gauge fields described by a simple probe Dirac-Born-Infeld (DBI) action we will demonstrate that
\beq \Phi = 2 \frac{\theta}{d} \eeq
where $d$ is the number of spatial dimensions. We will also see the need to allow a different HV exponent in the matter sector, $\theta_m$, as in the probe limit the probe and background energy densities can (and do) have a different scaling dimension. Different values for $\Phi$ can be realized in theories coupled to a dilaton field. We will in particular show that D$d$/D$q$ intersections give a string theoretic realization of this latter scenario.
Results similar to what we report here have appeared previously in \cite{Gouteraux:2014hca,Gouteraux:2013oca,Edalati:2013tma,Gath:2012pg,Gouteraux:2012yr,Gouteraux:2011ce,Charmousis:2010zz}. Especially in the more recent papers by Gouteraux  and Gouteraux and Kiritsis the need for a separate scaling exponent governing the gauge fields (our $\Phi$ above) had already been emphasized. We hope by succinctly summarizing the properties of these systems in terms of simple field theory scaling laws we can add some clarity to the discussion.

In \cite{Hartnoll:2009ns} it was demonstrated that the in the special case of $\theta=0$ the scaling laws imply that for $z=2$ the DC resistivity grows linearly with temperature, raising the hopes that this system may be related to the enigmatic strange metal phase of high-T$_c$ superconductors. In the same work it was however also found that the very same scaling implies a temperature dependence of the Hall angle that is inconsistent with what is seen in strange metals. We will demonstrate that the more general scaling laws described in here allow for the observed strange metal scaling of both conductivities. Our scaling also makes firm predictions for the dimensions of all thermo-electric coefficients, so it would be curious to explore to what extend those can be matched against the properties of strange metals.

The organization of this note is as follows: in the next section we review and determine holographically properties of systems based on HV metrics using the probe brane approach to conductivities \cite{Karch:2007pd,O'Bannon:2007in}.
In section 3 we will use the bulk form of the action to derive the correct scaling properties of the dual field theory and show that they reproduce all the data-points collected in section 2. In section 4 we discuss the D$d$/D$q$ system as an example of $\Phi \neq 2 \theta/d$; we once more can show perfect matching between bulk and boundary calculations. In Section 5 we briefly comment on similar results that had been obtained previously for the Einstein-Maxwell-Dilaton system and show that they nicely fit into our framework. In section 6 we conclude with a discussion of the putative connection to strange metals.

\section{``Experimental" Data Points}

In this section we collect a few facts about the thermodynamics and transport properties of bulk theories based on HV metrics, as well as charged matter described by a simple DBI probe brane\footnote{There is some ambiguity how to include coupling to background scalars in the DBI action. In this section we simply take the background geometry as a given and put a purely geometric DBI action ${\cal L} \sim \sqrt{g+F}$ on this background. In the earlier analysis of \cite{Charmousis:2010zz} the background dilaton already appeared in the DBI action, giving rise to different scaling relations. We will discuss the case of a non-trivial dilaton coupling to DBI in our section on the D$d$/D$q$ system. For now the main purpose of our analysis is to find the simplest holographic system that allows us to extract the pattern of scaling dimensions in the dual field theory. The set-up is also different from our earlier study of HV probe systems \cite{Ammon:2012je} where the HV theory was set up by the probe fields themselves instead of the background geometry.}.
We can write a metric which allows for a non-trivial scale symmetry with dynamical critical exponent $z$ and hyperscaling violating exponent $\theta$ in the form
\beq \label{hvmetric} g_{tt} = r^{\alpha}, \quad  g_{xx} = r^{\Delta}, \quad g_{rr}=r^{\beta}.\eeq
By reparametrizing $r$ this metric has been cast in different forms in the literature, but the coordinate invariant statements we would like to incorporate are that the scaling transformation gives
\beq x \rightarrow \lambda x, \quad t \rightarrow \lambda^z t, \quad ds \rightarrow \lambda^{\theta/d} ds.\eeq
Under this scaling $r \rightarrow \lambda^{\delta} r$ and hence
\beq \label{relation} \Delta \delta = \frac{2 \theta}{d} -2, \quad   \delta \alpha = \frac{2 \theta}{d} - 2 z.\eeq
While these relations are sufficient for our calculations, one convenient coordinate choice for this metric we'll sometimes use is
\beq \label{preferred} g_{tt}=-r^{2(\theta/d-z)}, \quad g_{rr}=g_{xx} = r^{ 2 (\theta/d-1)}. \eeq
This is the form used, for example, in \cite{Dong:2012se}.

At finite temperature we need to modify the HV metric to include a horizon at a horizon radius $r_h$. One family of such solutions has been found in \cite{Gouteraux:2011ce,Charmousis:2010zz,Huijse:2011ef}, where we will mostly follow the last reference. They use a metric with $\Delta=-2$ (that is $g_{xx} \sim r^{-2})$). In this coordinate system they find $r_h \sim T^{\frac{-(d- \theta)}{dz}}$ which in our preferred coordinates of \eqref{preferred}, where $g_{rr} \sim g_{xx}$ near the asymptotic boundary, implies
\beq r_h \sim T^{-1/z}. \eeq

\subsection{Thermodynamics and Dispersion Relation}

The defining relations of a $d$ spatial dimensional system with dynamical critical exponent $z$ and hyperscaling violating exponent $\theta$ are that the zero temperature dispersion relation is given by
\beq \omega = k^z \eeq
and the entropy density scales as
\beq s \sim T^{\frac{d-\theta}{z}}. \eeq
These are indeed both properties of the background geometry \eqref{hvmetric} and its finite temperature extension \cite{Huijse:2011ef}.

\subsection{DC Conductivities at Zero Temperature}
\label{sec:introduction}

In order to calculate conductivities we need to add a gauge field to the bulk action. Calculations are particularly easy if we use a bulk gauge field described by a probe brane, that is we use the DBI action for the gauge field and ignore its backreaction on the spacetime geometry.
In a spacetime with metric
$$ ds^2=g_{tt} dt^2 + g_{rr} dr^2 + g_{xx} d\vec{x}^2$$
where $d \vec{x}^2$ is a $d$ spatial dimensional flat space and all metric functions only depend on the radial direction the DBI action gives a non-linear DC conductivity \cite{Karch:2007pd}
\beq \sigma = \sqrt{A^2 g_{xx}^{d_s-2} + g_{xx}^{-2} n^2} \eeq
$A$ is a normalization constant that depends on the prefactor of the DBI and $n$ is the appropriately normalized charge density. $d_s$ is the number of spatial dimensions occupied by the probe matter fields, which can be less than $d$: $d_s \leq d$. The metric functions are to be evaluated at a critical position $r_*$ which is determined by
\beq g_{tt} g_{xx} = E^2 .  \label{rstar} \eeq

Already at zero temperature there are various limits in which the conductivity takes a simple scaling form as we only turn on one dimensionful quantity. The first scenario is to work at zero density, $n=0$. In this limit we only get a scale from the electric field itself, so
\beq \sigma = A g_{xx}^{(d_s-2)/2} \sim E^{\gamma}. \eeq
These non-linear conductivities, together with a microscopic model of how they could arise, where discussed in detail in \cite{Karch:2010kt}.

Alternatively, we can study the system in the limit of large density where the second term dominates. As in \cite{Hartnoll:2009ns} we can exploit the fact that in this limit the DBI conductivity is linear in density to find a scaling form:
\beq \label{densitysigma} \sigma = n g_{xx}^{-1} \sim n E^{\tilde{\gamma}}.\eeq
At first sight this finite conductivity in a translationally invariant system at finite charge density should be impossible. Without a mechanism to dissipate momentum the charge carriers should be accelerated by the external field without limit. The steady state described by \eqref{densitysigma} owes its existence to the probe limit. The gravitational bulk describes a neutral heat bath with a large number of degrees of freedom, $N^2$ if the dual is a large $N$ gauge theory. There are only order $N$ charged degrees of freedom which dissipate their momentum to the heat bath. Over time of course this dissipation will heat up the background heat bath, but in the large $N$ limit this backreaction can be ignored at least as long as one doesn't let the system evolve for times that are parametrically long in $N$.

The relevant information about the conductivity is encoded in the two scaling exponents $\gamma$ and $\tilde{\gamma}$. From \eqref{rstar} we get
\beq
r_*^{\Delta+\alpha} = E^2
\eeq
which implies that, at $r=r_*$, we have
\beq
g_{xx} = r_*^{\Delta} = E^{\frac{2 \Delta}{\Delta + \alpha}} = E^{ \frac{2(\theta-d)}{2 \theta - (z+1) d}}.
\eeq
Reassuringly $\delta$, our choice of radial variable, dropped out from the relation. The conductivity was smart enough not to pay attention to a choice of coordinate system. With this we can read of our two exponents
\beq
\label{gammas}
\gamma = (d_s-2) \frac{(\theta-d)}{2 \theta - (z+1) d}, \quad \quad \tilde{\gamma}
= \frac{2(d-\theta)}{2 \theta - (z+1) d}.
\eeq

\subsection{Hall Conductivity}

If we include finite magnetic fields, many more opportunities arise. The full non-linear conductivities can be mapped out similar to above \cite{O'Bannon:2007in}. One universal relation one finds at zero temperature in the small electric field limit is
\beq \sigma_{xy} = \frac{n}{B} \eeq
independent of $d$, $d_s$, $\theta$ and $z$.

\subsection{DC Conductivity at Finite Temperature}

To get some meaningful scaling relations at finite temperature it is best to look at the case of small electric fields, so that $E$ does not set an additional scale the conductivities can depend on. For small $E$ we have from \eqref{rstar} that $g_{tt} g_{xx}|_{r=r_*} \sim 0$ and so
\beq
r_* = r_h
\eeq
since $g_{tt}(r_h)=0$ and $g_{xx}$ is non-vanishing everywhere. In this case we can again look at the two contributions to the conductivity separately. At zero density we have
$\sigma \sim T^{\kappa}$, at large density we have conductivities proportional to density $\sigma \sim n T^{\tilde{\kappa}}$ with
\beq
\label{kappas}
\kappa = (d_s-2) \frac{ (d-\theta)}{d z}, \quad \tilde{\kappa} = \frac{2 (\theta-d)}{d z}.
\eeq
Last but not least let as look at the finite temperature Hall conductivity. The general expression is \cite{O'Bannon:2007in}
\beq
\sigma_{xy} = \frac{n B}{g_{xx}^2 + B^2}.
\eeq
The presence of non-zero $B$ also modifies the expression for $r_*$, but for small fields we can still set $r_*=r_h$ (temperature is the dominant scale). In this case we get
\beq
\label{finitehall}
\sigma_{xy} = n B g_{xx}^{-2} \sim n B T^{\frac{4 (\theta-d)}{d z}}.
\eeq

\subsection{Probe Brane Thermodynamics}

One interesting aspect of the probe brane system is that we have a clear separation between bulk fields and probe fields. As such we can define two separate free energy densities, the bulk and the probe free energy density, which do not have to have the same scaling behavior. This is most obvious in the case where $d_s<d$. In this case the two contributions to the free energy density can be distinguished by living at different places in space. Even in the case without any HV, they have different dimension, $d+z$ and $d_s+z$ respectively. In the presence of HV, where the scaling dimension of the bulk free energy density changes to $d+z-\theta$, the most naive expectation would be for the probe free energy to change to $d+z-\theta \frac{d_s}{d}$. That is each spatial direction picks up the same share of the anomalous scaling. For $d_s<d$ the net contribution of $\theta$ gets simply prorated. In order to confirm that this is born out for the probe, we need to calculate the free energy density of the probe.

The thermodynamics of simple DBI probes in HV geometries has been analyzed in \cite{Dey:2013vja}. At zero density the on-shell action of the brane is given by its area. This gives rise to a matter free energy (in the coordinates of \eqref{preferred})
\beq
\label{thermoone}
f_m \sim \int_{r_h} dr \,  r^{-t_1} \sim T^{\frac{t_1 -1}{z}}, \quad t_1 =d_s +z +1- \frac{(d_s+2)}{d} \theta .\eeq
To study the system at finite density, we need to include the $A_t$ component of the gauge field in the action. The DBI Lagrangian becomes
\beq {\cal L} \sim  r^{-t_1} \sqrt{1 - r^{2+2z-4 \frac{\theta}{d}} (A_t')^2}, \quad \quad
n = \frac{\delta {\cal L}}{\delta A_t'}. \eeq
From this we can easily get the scaling of matter free energy and chemical potential at zero temperature:
\beq
\label{thermotwo}
f_m \sim \int dr \, {\cal L} \sim n^{\frac{t_1-1}{d_s - \theta \frac{d_s}{d}}}, \quad \mu \sim \int dr \, A_t' \sim
n^{\frac{z-\frac{2 \theta}{d}}{d_s-\theta \frac{d_s}{d}}}.
\eeq

Note that even if $d_s=d$ we can meaningfully separate the two free energies by their scaling in $N$. In a large $N$ gauge theory we can have a contribution to the free energy density scaling as $N^2$ (the bulk) and one scaling as $N$ (the probe). They are clearly separate entities and can scale differently with temperature. At finite $N$ the system only has a single free energy and unless all contributions to the free energy have the same scale dimension of course scaling will no longer be a good symmetry. So in a system with $d_s=d$ with a matter free energy scaling differently from the bulk free energy, scaling is only an approximate symmetry emerging in the large $N$ limit.

\section{Holographic Scaling Relations.}
\label{scalinglaws}

We have seen that the holographic calculations from the previous sections give us simple scaling forms for, among other quantities, the conductivity whenever we are in the regime where it depends only on a single dimensionful quantity. Naively one would have expected these scaling relations to be given by assigning field theory dimensions according to \eqref{naturalscalings}. Unfortunately these natural dimension assignments predict
\beq
\gamma =\frac{d-\theta -2}{1+z}, \quad \tilde{\gamma} = \frac{-2}{1+z}
\eeq
which are much simpler than what the DBI finds and in gross contradiction to it.

The key to derive the correct scaling laws in the dual field theory is to realize that for a HV metric scaling is in fact not an isometry in the bulk. This distinguishes HV metrics from their $\theta=0$ cousins. Under a scale transformation $x \rightarrow \lambda x$ the metric scales by an overall prefactor
\beq ds \rightarrow \lambda^{\frac{\theta}{d}} \, ds. \eeq
With this the bulk action is actually not invariant, but scales with an overall prefactor as well. A transformation under which the action changes by an overall prefactor still gives rise to a symmetry of the classical equations of motion. Of course the on-shell action itself is no longer invariant, but picks up an extra scaling term which is responsible for the HV behavior of the free energy \eqref{free}.

The fact that the metric is not invariant under scaling however puts strong restrictions on the matter we can add to the theory without spoiling the invariance of the classical equation of motions. Every new term we add to the action has to scale with the same overall prefactor. This requires that we have to assign separate scaling properties to the various matter fields, that is the field theory operators and sources. For a gauge field described by a DBI action this means in particular that $F_{MN} F^{MN}$ has to have the same scaling as 1 so that $\sqrt{g} F^2$ and $\sqrt{g}$ scale with the same overall prefactor.

With this insight, we can deduce the following scaling laws. In the gravitational sector, we use the natural (defining) scaling properties:
\beq [x]=-1, \quad [t]=-z,  \quad [T]=z, \quad [f] = d+z-\theta.\eeq
This ensures that we get the correct behavior for the entropy density and dispersion:
\beq [\omega]=z, \quad [k]=1, \quad [s] = [f]-[T] = (d-\theta) \quad \Rightarrow \quad s\sim T^{\frac{d-\theta}{z}} . \eeq
In the gauge field sector however we need to assign anomalous scaling properties to $A_t$ (and hence the chemical potential) and $A_i$, that is they no longer have the same dimension as $t^{-1}$ and $x^{-1}$ respectively.
\beq [A_i]=1- \Phi, \quad [A_0]=z-\Phi=[\mu], \quad [E]=1+z-\Phi, \quad [B]=2-\Phi .\eeq
For the DBI action we have
\beq \Phi = \frac{2 \theta}{d}. \eeq
That is each $A$ cancels the transformation property of one inverse metric in $F^2$. Usually the scaling dimensions of $A_i$ and $x$ as well as $A_0$ and $t$ are linked with each other by gauge invariance, as they appear together in the covariant derivative. They can be divorced by a dimensionful coupling constant appearing in front of the gauge field in the covariant derivative.

Thermodynamic quantities for the gauge field sector introduce one additional free HV exponent
\beq \quad [n]=d_s-\theta_m.\eeq
For the DBI probe brane we have
\beq
\theta_m = \frac{d_s}{d} \theta.
\eeq
We will see holographic examples with different values of $\theta_m$ in the next section.
Note that in the case where $d_s<d$, that is the field theory currents are localized on a submanifold that is not filling all of space, $n$ (and hence the matter free energy density with $[f_m]=[n]+[\mu]$) picks up only a fraction $d_s/d$ of the anomalous scaling implied by $\theta$. In the bulk this can easily be seen from the fact the determinant of the induced metric simply picks up fewer metric factors than the determinant of the full spacetime metric.

Conservation laws now fix the dimension of the current:
\beq
[n]=d_s - \theta_m \quad \Rightarrow \quad [j]=d_s - \theta_m + z -1.\eeq
For the DBI probe this evaluates to
\beq
[j]=d_s+z-1 -\theta \frac{d_s}{d}.
\eeq
We can easily deduce from this the scaling of the matter free energy
\beq
[f_m]=[\mu]+[n] =d_s+z - \Phi - \theta_m. \label{freescaling}
\eeq
It is easy to check that for the DBI probe this scaling perfectly reproduces our holographic calculations \eqref{thermoone} and \eqref{thermotwo}.

Last but not least, let us match the results for the conductivity. The dimensions of the conductivity (and conductivity divided by particle density) are
\beq
\label{sigmascaling}
[\sigma] = [j] - [E] = d_s + \Phi- \theta_m - 2, \quad [\frac{\sigma}{n}] = \Phi-2
\eeq
For the DBI probe brane this evaluates to
\beq
[\sigma]=
 \frac{d_s-2}{d} (d-\theta), \quad
[\frac{\sigma}{n}] = 2 \frac{\theta-d}{d}.
\eeq
This scaling perfectly reproduces the zero temperature exponents from \eqref{gammas} as well as the finite temperature exponents from \eqref{kappas}.
Note that with our assignments we also get the zero $T$ Hall conductivity right, since
\beq [n] - [B] = \frac{d-\theta}{d} (d_s-2) = [\sigma] .\eeq
It is also easy to see that this scaling perfectly reproduces the finite $T$ Hall conductivity \eqref{finitehall} as well. Note that for any choice of $\Phi$ and $\theta_m$ we have
\beq
\label{noway}
[\frac{\sigma}{n B} ] = 2 \Phi - 4 = 2 [\frac{\sigma}{n}].
\eeq
This relation will become important for us later when we discuss potential applications to strange metals.

As we discussed before, for generic matter HV exponents $\Phi$ and $\theta_m$ the free energy of the matter sector has a different dimension from the free energy in the bulk sector. This is consistent only if $d_s <d$, so we can physically separate the localized energy density from the bulk energy density, or when the matter is a probe field, so we can physically separate the order $N^2$ energy density of the bulk from the order $N$ density of the matter fields. In both those case, $f$ and $f_m$ are independent physical quantities which have their own dimension under scaling. If we are unable to separate the free energy into a bulk and a matter contribution, our system only has scaling as a symmetry if we can assign a unique dimension to the full free energy, that is when $[f]=[f_m]$, or in other words
\beq
\label{samef}
[f] = [f_m] \quad \Leftrightarrow \quad \theta = \theta_m + \Phi.
\eeq
In the last section of this paper we will discuss potential phenomenological applications of our scaling laws. In that case we will always insist on \eqref{samef} being obeyed, leaving $\Phi$ as the only new scaling parameter. But
since all our ``experimental" scaling facts have been obtained for DBI probe branes we for now can (and have to) use $\theta_m$ and $\Phi$ as independent parameters.

\section{The Dd/Dq system}

As we have seen in the previous section, in systems with HV violating metrics the scaling properties of the dual field theory are not uniquely determined by the gravitational background, but also depend on the {\it action} we chose for the matter fields. The general rule is that we need to assign the matter fields anomalous transformation properties so that the matter action scales with a universal prefactor (giving rise to $\theta_m$). Of course it may not always be possible to find any scaling rule for the matter fields in which this is true. In this case scaling is not a good symmetry of the system with the matter sources turned on (or for correlation functions involving the matter fields). For DBI we needed to assign the vector potential $A_{\mu}$ transformation properties that cancel that of a single inverse metric, giving rise to $\Phi=2 \theta/d$. A simple example for a gauge field action that has scaling as a symmetry but with a different $\Phi$ is given by a Maxwell-Dilaton system with Lagrangian
\beq {\cal L} \sim \sqrt{g} \left ( 1 + e^{- \Psi} F^2 + \ldots \right ) .\eeq
In a non-trivial dilaton background the scaling of $A_{\mu}$ needs to compensate an inverse metric as well as the scaling $e^{-\Psi}$. With a dilaton, we can also get more interesting values for $\theta_m$. In this section we want to study a particular class of examples of this type where the scalings can easily be worked out along the lines above. The fact that we can reproduce the highly non-trivial bulk scalings in this case gives a great confirmation of our construction.

\subsection{Brief review of the Dd/Dq system}

A well studied string theory embedding of a system with an HV metric that also allows for probe branes non-trivially coupled with a dilaton prefactor is provided by the D$d$/D$q$ system. That is, we study a D$q$ probe in the background geometry of a D$d$ brane. The holographically dual field theory of these theories is well understood: maximally supersymmetric Yang-Mills theory in $d+1$ dimensions coupled to fundamental representation matter, whose details vary with how the D$q$ is embedded \cite{Itzhaki:1998dd,Mateos:2007vn}.
The full 9+1 dimensional D$d$ brane metric and dilaton background are given by
\beq ds^2 = H^{-1/2} (-dt^2 + dx_d^2) + H^{1/2} \left ( du^2 + u^2 d\Omega_{8-d}^2 \right ), \quad e^{\Psi}   = H^{\frac{3-d}{4}}. \eeq
In the near horizon limit we have (setting the curvature radius to 1)
\beq
H = u^{d-7}.
\eeq
The D$d$ brane allows a near-extremal generalization with a horizon radius $u_h$ that scales with temperature as
\beq u_h \sim T^{\frac{2}{5-d}}.\eeq
The entropy density of this near-extremal brane scales as
\beq
\label{dddqentro}
s \sim T^{\frac{9-d}{5-d}}.
\eeq
To see that this is indeed a HV scaling background, one wants to reduce on the internal $8-d$ sphere and go to the $d+2$ dimensional Einstein frame \cite{Dong:2012se}. Since the prefactor of the $d+2$ dimensional Einstein Hilbert term in the string frame is $e^{-2 \Phi} H^{\frac{8-d}{4}} u^{8-d} = H^{\frac{d+2}{4}} u^{8-d}$ the rescaling to Einstein frame is given by
\beq
g_{\mu \nu} \rightarrow g_{\mu \nu} H^{\frac{d+2}{2d}} u^{\frac{2 (8-d)}{d}} = g_{\mu \nu} u^{\frac{(d-3)(d-6)}{2d}} .
\eeq
The resulting Einstein frame metric
\beq
ds^2_{d+2} = u^{(16-2d)} H^{1/d} \left ( - dt^2 + dx_d^2 + H du^2 \right )
\eeq
can easily be seen to be of HV scaling form \cite{Dong:2012se} with
\beq
\label{thetadd}
\theta = d - \frac{9-d}{5-d} = - \frac{(d-3)^2}{5-d}, \quad \quad z=1
\eeq
This scaling is in perfect agreement with the thermodynamics \eqref{dddqentro} of the D$d$ system.
To reach our preferred coordinate system with $g_{xx}=g_{rr}$ we need to
change to a new radial coordinate that absorbs the prefactor $H$ in front of $u$:
\beq
r=\int \sqrt{H} du \sim u^{\frac{d-5}{2}}.
\eeq
In this coordinate system the horizon radius simply scales as $r_h \sim 1/T$.

For a D$q$ probe brane wrapping $q-d_s-1$ internal dimension, the $d_s+2$ dimensional volume and Maxwell terms on the brane worldvolume we get after reduction on the internal space and going to the Einstein frame are given by
\beq
\label{dddqaction}
{\cal L} = r^{a_1} \sqrt{g} \left ( 1+ r^{\zeta}  F_{\mu \nu} F^{\mu \nu} + \ldots \right )
\eeq
with
\beq
\label{zeta} \zeta = - \frac{2}{d-5} \frac{(d-3)(d-6)}{d}.
\eeq
$\zeta$ sets the anomalous scaling of the gauge field $\Phi$.
$a_1$ gives the overall scaling of the free energy and so will be responsible for $\theta_m$. But it is easier to calculate $f_m$ directly from the 10d metric as we'll do momentarily, so we will not worry about the value of $a_1$.

\subsection{``Experimental" facts}

The thermodynamics associated with the background geometry of the D$d$ brane has already been accounted for by the background $\theta$ of \eqref{thetadd}. What we are concerned with here are the properties of the probe D$q$ brane
extending over $d_s$ spatial dimensions of the $d$ dimensional field theory. Its free energy is given by the on-shell action, which is given by a DBI action multiplied with $e^{-\Psi}$. From this one finds that, at zero density, the free energy scales as  \cite{Mateos:2007vn,Karch:2009eb}
\beq \label{fdddq} f \sim T^{\frac{2}{5-d} (\nu+1)} \eeq
whereas at zero temperature, finite density it scales as
\beq
\label{factone}
f \sim n^{1+\frac{1}{\nu}}
\eeq
with
\beq
\nu = \frac{(d-7)(q-2 d_s - 4 +d)}{4} + q - d_s -1.
\eeq

The DC conductivities for this system have been worked out in \cite{Karch:2009eb}. One once more finds simple scaling laws at zero density as well as in the density dominated limit. They are given by
\beq
\sigma \sim  T^{\frac{2}{5-d} (\nu + \frac{d-7}{2} )}
\eeq
and
\beq
\label{factlast}
\sigma \sim n T^{\frac{d-7}{5-d}}
\eeq
respectively.

\subsection{Deriving the anomalous scaling factors}

For the gauge field in the action \eqref{dddqaction} to transform properly under scaling when compared to the area term, we need to assign the gauge fields the anomalous scaling exponent
\beq \label{phidddq}
\Phi = 2 \frac{\theta}{d} + \frac{ \zeta}{2} = \frac{d-3}{d-5}.
\eeq
On the other hand, the scaling of the free energy with temperature \eqref{fdddq} tells us that
\beq
[f_m] = \frac{2}{5-d} (\nu +1)
\eeq
This implies (according to \eqref{freescaling} and using $\Phi$ from \eqref{phidddq} above) a matter HV exponent
\beq
\theta_m = \frac{(d-3)(d+q-8)}{2 (d-5)}.
\eeq
It is straightforward to check that with these two assignments for the new HV exponents introduced in section \ref{scalinglaws}, our dimension assignments accurately reproduce the holographic results from eqs. \eqref{factone} through \eqref{factlast}.

\section{Einstein-Maxwell-Dilaton System}

A large class of HV solutions have been obtained in \cite{Gouteraux:2014hca,Gouteraux:2013oca,Edalati:2013tma,Gath:2012pg,Gouteraux:2012yr,Gouteraux:2011ce,Charmousis:2010zz}
based on the Einstein-Maxwell-Dilaton (EMD) system. In the EMD system the matter fields in the bulk are fully backreacted and so it is nice to compare and contrast these solutions to our analysis in the previous sections, which was mostly based on probe branes. Since in EMD there is no defect present in the field theory these theories work with $d_s=d$. For the system to have a scaling symmetry, all terms in the action need to scale the same in we need to enforce \eqref{samef}. Eliminating $\theta_m$ in terms of $\Phi$ this implies
\beq
\label{newsigmascaling}
[\sigma]  = d + 2 \Phi- \theta - 2 .
\eeq
Comparing in particular with \cite{Gouteraux:2014hca,Gouteraux:2013oca} we see that this scaling agrees with the behavior found in there as long as we identify our scaling exponent $\Phi$ with their $\zeta$ via
\beq
\Phi = \frac{\zeta+\theta-d}{2}.
\eeq
It is straightforward to verify that with this assignment indeed all terms in the EMD scale the same. The non-trivial scaling of the gauge field implied by $\Phi$ is forced upon us by the coupling to the dilaton. The gauge coupling in the bulk itself is dimensionful, but these dimensions can be completely accounted for by powers of the AdS curvature radius $L$ and do not correspond to a non-trivial $\Phi$. In order for the gauge coupling to scale non-trivially in the sense that $\Phi \neq 0$, we need it to be set by a scalar field that itself has a non-trivial radial profile. It is indeed easy to see that if we set the dilaton potential and gauge coupling to a constant, that is chose $\gamma=\delta=0$ in \cite{Gouteraux:2014hca,Gouteraux:2013oca}, we get
$\theta=0$, $\zeta=d$ and hence $\Phi=0$: scaling is actually an isometry in that case.

\section{Discussion: Connection to strange metals?}

In this work we have demonstrated that the electromagnetic properties of critical points with non-trivial HV are characterized by two novel exponents, $\Phi$ and $\theta_m$ introduced in section \ref{scalinglaws}, together with the familiar exponents $\theta$ and $z$. While we gave explicit holographic examples of theories realizing various values of these scaling parameters (and additional examples in a similar spirit can e.g. be found in \cite{Lee:2010uy,Lee:2010ez}), phenomenologically the most reasonable approach seems to be to just treat $\Phi$ and $\theta_m$ as free parameters characterizing any putative critical point and to then investigate whether such scaling can give rise to phenomenological acceptable predictions.

One system that one may try to apply our results to are strange metals. One of the enigmatic properties of these particular non-Fermi liquids is their resistivity's linear growth with temperature. In \cite{Hartnoll:2009ns} it was pointed out that this behavior could be reproduced from a scale invariant theory with non-trivial dynamical critical exponent $z$. According to \eqref{sigmascaling}, the dimension of $\sigma/n$ is $\Phi-2$. So if we assume our strange metal operates in the regime where the conductivity is linear in density, this tells us that already for $\Phi=0$ we get $\sigma^{-1} \sim T^{2/z}$ and so for $z=2$ one gets a linear grows of the resistivity \cite{Hartnoll:2009ns}. Allowing for the more general scaling studied in this paper, we see that linear resistivity of the density dominated conductivity arises whenever\footnote{The fact that hyperscaling violation can give rise to linear resistivity for $z=2-2 \frac{\theta}{d}$ is also already implicit in \cite{Dey:2013vja,Edalati:2012tc,Edalati:2013tma}.} $z=2-\Phi$. It was however also pointed out in \cite{Hartnoll:2009ns} that the very same scaling fails to reproduce the correct behavior of the Hall conductivity. In strange metals, the linear temperature rise in $\sigma_{xx}^{-1}$ comes along with a $T^3$ scaling of $\sigma_{xy}^{-1}$. Assuming that the Hall conductivity is both linear in density and magnetic field (as it is in holographic probe brane constructions) one can see from \eqref{noway} that the scaling with temperature of the Hall conductivity is always twice that of the density dominated $\sigma_{xx}$. So if the latter goes as $1/T$, the former scales as $1/T^2$, not $1/T^3$. The introduction of the new free parameter $\Phi$ did not help resolve this tension. The only way to reconcile the density dominated $\sigma_{xx}$ and $\sigma_{xy}$ with experiment is to chose different values of $\Phi$ for $A_i$ and $A_0$. One way to accomplish this is to start with a genuinely non-relativistic gravitational theory in the bulk as recently advertised in \cite{Griffin:2012qx,Janiszewski:2012nf,Janiszewski:2012nb}

Of course the other option is to operate the system in a regime where $\sigma_{xx}$ is not density dominated, but temperature dominated. The Hall conductivity still is linear in both density and magnetic field, so we will still have
\beq
\sigma_{xy} \sim B n T^{\frac{2 \Phi -4}{z}}.
\eeq
Since the dimension of $\sigma_{xx}$ itself according to \eqref{sigmascaling} is $d_s + \Phi - \theta_m -2$ we have
\beq
\label{xx}
\sigma_{xx} \sim T^{\frac{d_s + \Phi - \theta_m -2}{z}}.
\eeq
In this case, our new scaling exponents allow us to get the correct scalings for both conductivities. Getting $\sigma_{xy}$ to scale as $T^{-3}$ requires
\beq
\Phi=2 - \frac{3}{2} z.
\eeq
Using this in \eqref{xx} we see that $\sigma_{xx} \sim T^{\frac{d_s - \theta_m}{z} - \frac{3}{2}}$. This is inversely proportional to $T$ as long as
\beq
\theta_m = d_s - \frac{z}{2}.
\eeq
So the properties of the strange metal can easily be accommodated. If we in addition want to avoid $f_m$ and $f$ to have different scaling dimension, we can still impose \eqref{samef} by fixing $\theta$ itself to be
\beq
\theta = \theta_m + \Phi = d_s + 2 - 2z.
\eeq
As a concrete realization, we can e.g. chose a $d=d_s=2$ dimensional system with $z=2$ and chose $\Phi=-1$, $\theta_m=1$, $\theta=0$. Whether such a scaling can be obtained either from a holographic model or some microscopic Lagrangian will be left as an open question.

It should be noted that once our HV coefficients are fixed, there is no more freedom in setting the scaling laws for all thermoelectric coefficients. Dimensional analysis now can be applied to all thermoelectric phenomena, such as the Nernst and Seebeck effect. The scaling dimensions also constrain the frequency and wavenumber dependence of current correlation functions. This gives in principle many more data points one can look at in order to determine whether our generalized HV critical points can give a correct physical description of strange metals or any other quantum critical system.

\section*{Acknowledgments}
\noindent I'd like to thank Sean Hartnoll for very useful correspondence and encouragement. Thanks also to Blaise Gouteraux for helping to clarify the connection between the approach followed in here and his original work, as discussed in section 5. This work was supported, in part, by the US Department of Energy under grant number DE-FG02-96ER40956.

\bibliographystyle{JHEP}
\bibliography{hvscaling}

\providecommand{\href}[2]{#2}\begingroup\raggedright\begin{thebibliography}{10}

\bibitem{Maldacena:1997re}
J.~M. Maldacena, {\it {The Large N limit of superconformal field theories and
  supergravity}},  {\em Adv.Theor.Math.Phys.} {\bf 2} (1998) 231--252,
  \href{http://xxx.lanl.gov/abs/hep-th/9711200}{{\tt hep-th/9711200}}.

\bibitem{Witten:1998qj}
E.~Witten, {\it {Anti-de Sitter space and holography}},  {\em
  Adv.Theor.Math.Phys.} {\bf 2} (1998) 253--291,
  \href{http://xxx.lanl.gov/abs/hep-th/9802150}{{\tt hep-th/9802150}}.

\bibitem{Gubser:1998bc}
S.~Gubser, I.~R. Klebanov, and A.~M. Polyakov, {\it {Gauge theory correlators
  from noncritical string theory}},  {\em Phys.Lett.} {\bf B428} (1998)
  105--114, \href{http://xxx.lanl.gov/abs/hep-th/9802109}{{\tt
  hep-th/9802109}}.

\bibitem{Kachru:2008yh}
S.~Kachru, X.~Liu, and M.~Mulligan, {\it {Gravity duals of Lifshitz-like fixed
  points}},  {\em Phys.Rev.} {\bf D78} (2008) 106005,
  \href{http://xxx.lanl.gov/abs/0808.1725}{{\tt 0808.1725}}.

\bibitem{Ogawa:2011bz}
N.~Ogawa, T.~Takayanagi, and T.~Ugajin, {\it {Holographic Fermi Surfaces and
  Entanglement Entropy}},  {\em JHEP} {\bf 1201} (2012) 125,
  \href{http://xxx.lanl.gov/abs/1111.1023}{{\tt 1111.1023}}.

\bibitem{Huijse:2011ef}
L.~Huijse, S.~Sachdev, and B.~Swingle, {\it {Hidden Fermi surfaces in
  compressible states of gauge-gravity duality}},  {\em Phys.Rev.} {\bf B85}
  (2012) 035121, \href{http://xxx.lanl.gov/abs/1112.0573}{{\tt 1112.0573}}.

\bibitem{Dong:2012se}
X.~Dong, S.~Harrison, S.~Kachru, G.~Torroba, and H.~Wang, {\it {Aspects of
  holography for theories with hyperscaling violation}},  {\em JHEP} {\bf 1206}
  (2012) 041, \href{http://xxx.lanl.gov/abs/1201.1905}{{\tt 1201.1905}}.

\bibitem{Hartnoll:2009ns}
S.~A. Hartnoll, J.~Polchinski, E.~Silverstein, and D.~Tong, {\it {Towards
  strange metallic holography}},  {\em JHEP} {\bf 1004} (2010) 120,
  \href{http://xxx.lanl.gov/abs/0912.1061}{{\tt 0912.1061}}.

\bibitem{Khveshchenko:2014nka}
D.~Khveshchenko, {\it {Taking a critical look at holographic critical matter}},
   \href{http://xxx.lanl.gov/abs/1404.7000}{{\tt 1404.7000}}.

\bibitem{Gouteraux:2014hca}
B.~Goutéraux, {\it {Charge transport in holography with momentum dissipation}},
   \href{http://xxx.lanl.gov/abs/1401.5436}{{\tt 1401.5436}}.

\bibitem{Gouteraux:2013oca}
B.~Goutéraux, {\it {Universal scaling properties of extremal cohesive
  holographic phases}},  {\em JHEP} {\bf 1401} (2014) 080,
  \href{http://xxx.lanl.gov/abs/1308.2084}{{\tt 1308.2084}}.

\bibitem{Edalati:2013tma}
M.~Edalati and J.~F. Pedraza, {\it {Aspects of Current Correlators in
  Holographic Theories with Hyperscaling Violation}},  {\em Phys.Rev.} {\bf
  D88} (2013) 086004, \href{http://xxx.lanl.gov/abs/1307.0808}{{\tt
  1307.0808}}.

\bibitem{Gath:2012pg}
J.~Gath, J.~Hartong, R.~Monteiro, and N.~A. Obers, {\it {Holographic Models for
  Theories with Hyperscaling Violation}},  {\em JHEP} {\bf 1304} (2013) 159,
  \href{http://xxx.lanl.gov/abs/1212.3263}{{\tt 1212.3263}}.

\bibitem{Gouteraux:2012yr}
B.~Gouteraux and E.~Kiritsis, {\it {Quantum critical lines in holographic
  phases with (un)broken symmetry}},  {\em JHEP} {\bf 1304} (2013) 053,
  \href{http://xxx.lanl.gov/abs/1212.2625}{{\tt 1212.2625}}.

\bibitem{Gouteraux:2011ce}
B.~Gouteraux and E.~Kiritsis, {\it {Generalized Holographic Quantum Criticality
  at Finite Density}},  {\em JHEP} {\bf 1112} (2011) 036,
  \href{http://xxx.lanl.gov/abs/1107.2116}{{\tt 1107.2116}}.

\bibitem{Charmousis:2010zz}
C.~Charmousis, B.~Gouteraux, B.~S. Kim, E.~Kiritsis, and R.~Meyer, {\it
  {Effective Holographic Theories for low-temperature condensed matter
  systems}},  {\em JHEP} {\bf 1011} (2010) 151,
  \href{http://xxx.lanl.gov/abs/1005.4690}{{\tt 1005.4690}}.

\bibitem{Karch:2007pd}
A.~Karch and A.~O'Bannon, {\it {Metallic AdS/CFT}},  {\em JHEP} {\bf 0709}
  (2007) 024, \href{http://xxx.lanl.gov/abs/0705.3870}{{\tt 0705.3870}}.

\bibitem{O'Bannon:2007in}
A.~O'Bannon, {\it {Hall Conductivity of Flavor Fields from AdS/CFT}},  {\em
  Phys.Rev.} {\bf D76} (2007) 086007,
  \href{http://xxx.lanl.gov/abs/0708.1994}{{\tt 0708.1994}}.

\bibitem{Ammon:2012je}
M.~Ammon, M.~Kaminski, and A.~Karch, {\it {Hyperscaling-Violation on Probe
  D-Branes}},  {\em JHEP} {\bf 1211} (2012) 028,
  \href{http://xxx.lanl.gov/abs/1207.1726}{{\tt 1207.1726}}.

\bibitem{Karch:2010kt}
A.~Karch and S.~Sondhi, {\it {Non-linear, Finite Frequency Quantum Critical
  Transport from AdS/CFT}},  {\em JHEP} {\bf 1101} (2011) 149,
  \href{http://xxx.lanl.gov/abs/1008.4134}{{\tt 1008.4134}}.

\bibitem{Dey:2013vja}
P.~Dey and S.~Roy, {\it {Zero sound in strange metals with hyperscaling
  violation from holography}},  {\em Phys.Rev.} {\bf D88} (2013) 046010,
  \href{http://xxx.lanl.gov/abs/1307.0195}{{\tt 1307.0195}}.

\bibitem{Itzhaki:1998dd}
N.~Itzhaki, J.~M. Maldacena, J.~Sonnenschein, and S.~Yankielowicz, {\it
  {Supergravity and the large N limit of theories with sixteen supercharges}},
  {\em Phys.Rev.} {\bf D58} (1998) 046004,
  \href{http://xxx.lanl.gov/abs/hep-th/9802042}{{\tt hep-th/9802042}}.

\bibitem{Mateos:2007vn}
D.~Mateos, R.~C. Myers, and R.~M. Thomson, {\it {Thermodynamics of the brane}},
   {\em JHEP} {\bf 0705} (2007) 067,
  \href{http://xxx.lanl.gov/abs/hep-th/0701132}{{\tt hep-th/0701132}}.

\bibitem{Karch:2009eb}
A.~Karch, M.~Kulaxizi, and A.~Parnachev, {\it {Notes on Properties of
  Holographic Matter}},  {\em JHEP} {\bf 0911} (2009) 017,
  \href{http://xxx.lanl.gov/abs/0908.3493}{{\tt 0908.3493}}.

\bibitem{Lee:2010uy}
B.-H. Lee and D.-W. Pang, {\it {Notes on Properties of Holographic Strange
  Metals}},  {\em Phys.Rev.} {\bf D82} (2010) 104011,
  \href{http://xxx.lanl.gov/abs/1006.4915}{{\tt 1006.4915}}.

\bibitem{Lee:2010ez}
B.-H. Lee, D.-W. Pang, and C.~Park, {\it {Zero Sound in Effective Holographic
  Theories}},  {\em JHEP} {\bf 1011} (2010) 120,
  \href{http://xxx.lanl.gov/abs/1009.3966}{{\tt 1009.3966}}.

\bibitem{Edalati:2012tc}
M.~Edalati, J.~F. Pedraza, and W.~Tangarife~Garcia, {\it {Quantum Fluctuations
  in Holographic Theories with Hyperscaling Violation}},  {\em Phys.Rev.} {\bf
  D87} (2013), no.~4 046001, \href{http://xxx.lanl.gov/abs/1210.6993}{{\tt
  1210.6993}}.

\bibitem{Griffin:2012qx}
T.~Griffin, P.~Horava, and C.~M. Melby-Thompson, {\it {Lifshitz Gravity for
  Lifshitz Holography}},  {\em Phys.Rev.Lett.} {\bf 110} (2013), no.~8 081602,
  \href{http://xxx.lanl.gov/abs/1211.4872}{{\tt 1211.4872}}.

\bibitem{Janiszewski:2012nf}
S.~Janiszewski and A.~Karch, {\it {String Theory Embeddings of Nonrelativistic
  Field Theories and Their Holographic Horava Gravity Duals}},  {\em
  Phys.Rev.Lett.} {\bf 110} (2013), no.~8 081601,
  \href{http://xxx.lanl.gov/abs/1211.0010}{{\tt 1211.0010}}.

\bibitem{Janiszewski:2012nb}
S.~Janiszewski and A.~Karch, {\it {Non-relativistic holography from Horava
  gravity}},  {\em JHEP} {\bf 1302} (2013) 123,
  \href{http://xxx.lanl.gov/abs/1211.0005}{{\tt 1211.0005}}.

\end{thebibliography}\endgroup

\end{document}